\documentclass [extra,referee]{gji}
\usepackage{graphicx}
\usepackage{amsmath}

\begin{document}

\title{Frequency-dependent streaming potentials: a review}

\author[L. Jouniaux and C. Bordes ]
   { L. Jouniaux$^1$ and C. Bordes$^2$\\
 $^{1}$Institut de Physique du Globe de Strasbourg, UdS-CNRS UMR 7516 \\Universit\'e de Strasbourg,
5 rue Ren\'e Descartes, 67084 Strasbourg, FRANCE\\
  $^{2}$ Laboratoire des fluides complexes et de leurs r\'eservoirs, UPPA-CNRS UMR5150 \\ Universit\'e de Pau et des Pays de l'Adour, Pau, FRANCE\\}

\date{}
\maketitle

\begin{abstract}
This paper has been published in: International J. of Geophysics, vol 2012, article ID 648781, doi:10.1155/2012/648781 \\
open access at: www.hindawi.com/journals/ijgp/2012/648781 \\

The interpretation of seismoelectric observations involves the dynamic electrokinetic coupling, which is related to the streaming potential coefficient. We describe the different models of the frequency-dependent streaming potential, mainly the Packard's and the Pride's model. We compare the transition frequency separating low-frequency viscous flow and high-frequency inertial flow, for dynamic permeability and dynamic streaming potential. We show that the transition frequency, on a various collection of samples for which both formation factor and permeability are measured, is predicted to depend on the permeability as inversely proportional to the permeability. We review the experimental setups built to be able to perform dynamic measurements. And we present some measurements and calculations of the dynamic streaming potential.

\end{abstract}

\section{Introduction}

Electrokinetics arise from the interaction between the rock matrix and the pore water. Therefore electrokinetic
 phenomena are often observed in aquifers, volcanoes, and hydrocarbon or hydrothermal reservoirs. Observations
 show that seismoelectromagnetic signals associated to earthquakes can be induced by electromagnetic induction
 \citep{honkura09,matsus02} or by electrokinetic effect \citep{takeuchi98,fenoglio95}. 
The electrokinetic phenomena are due to pore pressure gradients leading to fluid flow in the porous media or 
fractures, and inducing 
electrical fields. These electrokinetic effects are associated to the electrical double layer which was
 originally described by Stern. The electrokinetic signals can be induced by global displacements of the 
reservoir fluids (streaming potential) or by the propagation of seismic waves (seismoelectromagnetic effect). 
As soon as these pressure gradients have a transient signature, the dynamic part of the electrokinetic coupling
 has to be taken into account by introducing the dependence on fluid transport properties. 


It is generally admitted that two kinds of seismoelectromagnetic effects can be observed. The dominant
 contribution, commonly called ``coseismic'', is generated close to the receivers during the passage of seismic
 waves. The second kind, so called ``interfacial conversion''\citep{dupuis09}, is very similar to dipole 
radiation and is generated at physico-chemical interfaces due to strong electrokinetic coupling 
discontinuities. This interface conversion is often perceived to have the potential to detect fine fluids
 transitions with higher resolution than seismic investigations, but in practice, signals are often masked 
by electromagnetic disturbances, especially when generated at great depth. 

Nevertheless recent field studies have focused on the seismo-electric conversions linked to electrokinetics 
in order to investigate oil and gas reservoirs \citep{thompson05} or hydraulic reservoirs
 \citep{dupuis06, dupuis07,dupuis09,strahser07,haines07a,haines07b,strahser11,garamb01}. It has been shown 
using these investigations that not only the depth of the reservoir can be deduced, but also the geometry of
 the reservoir can be imaged using the amplitudes of the electro-seismic signals \citep{thompson07}. Moreover
 fractured zones can be detected and permeability can be measured using seismo-electrics in borehole
 \citep{singer05,pain05,mikha00,jouniaux11b}. This method is especially appealing to hydrogeophysics for the 
detection of subsurface interfaces induced by contrasts in permeability, in porosity, or in electrical 
properties (salinity and water content) \citep{schakel11,schakel10,garamb02b}.

The analytical interpretation of the seismoelectromagnetic phenomenon has been described by \cite{pride94},
 by connecting the theory of \cite{biot56a} for the seismic wave propagation in a two phases medium with 
Maxwell's equations, using dynamic electrokinetic couplings. The seismoelectromagnetic conversions have been 
modeled in homogeneous or layered saturated media \citep{haar97,haar98,garamb01,garamb02b,gao10} with 
applications to reservoir geophysics \citep{saunders06}.

Theoretical developments showed that the electrical field induced by the $P$-waves propagation is related to 
the acceleration \citep{garamb01}. The electrokinetic coupling is created at the interface between grains and 
water, when there is a relative motion of electrolyte ions with respect to the mineral surface. Thus, seismic
 wave propagation in fluid-filled porous media generates conversions from seismic to electromagnetic energy 
which can be observed at the macroscopic scale, due to this electrokinetic coupling at the pore scale. The 
seismoelectric coupling is directly dependent on the fluid conductivity, 
the fluid density and the electric double-layer (the electrical interface between the grains and the water)
(see the tutorial by \citep{jouniaux11t}, in this special issue ``Electrokinetics in Earth Sciences' for more details). 
For more details on the surface complexation reactions see \cite{davis78}
 or \cite{guichet06}. It can be accurately quantified in the broad band by a dynamic coupling \citep{pride94} 
which can be linked in the low frequency limit to the steady-state streaming potential coefficient largely studied
  in porous media \citep{ishido81,pozzi94,joun95a,joun95b,joun97,joun94,joun99,joun00,guichet03,guichet06,jaafar09,jouniaux09,vinogradov10,jackson10,allegre10}.

Laboratory experiments have also been investigated for a better understanding of the seismoelectric conversions
\citep{migunov77,chandler81,mironov94,jiang98,zhu99,zhu00,zhu03,chen05,bordes06,block06,zhu08,bordes08}.
These papers describe the laboratory studies performed to investigate this dynamic coupling. An oscillating 
pore pessure must be applied to a rock sample, and because of the relative motion between the rock and the 
fluid, an induced streaming potential can be measured. Depending on the oscillating frequency of the fluid,
 the fluid makes a transition from viscous dominated flow to inertial dominated flow. As the frequency 
increases, the motion of the fluid within the rock is delayed and larger pressure is needed. In order to 
know the dynamic coupling, both real and imaginary part of the streaming potential must be measured.

\section{From dynamic streaming potential to seismoelectromagnetic coupling}

The steady-state streaming potential coefficient is defined as the ratio of the streaming potential to the driving 
pore pressure:

\begin{equation}
\mathbf{C_{s0}}= \frac{\Delta V}{\Delta P}= \frac{\epsilon \zeta} {\eta \sigma_{f}} 
	\label{eqn:Cs}
\end{equation}

which is called the Helmholtz-Smoluchowski equation, where $\sigma_{f}$, $\epsilon $ and $\eta$ are the fluid 
conductivity, the dielectric constant of the fluid, and the fluid dynamic viscosity respectively 
(see the tutorial by \citep{jouniaux11t}). In this formula the surface electrical conductivity is neglected compared to the fluid electrical conductivity. The potential $\zeta$  is the electrical potential within the double-layer on the slipping plane. Although the zeta potential can hardly be modeled for a rock and although it can not be direclty measured within a rock, the steady-state streaming potential coefficient can be measured in laboratory, by applying a fluid pressure difference ($\Delta P $) and by measuring the induced streaming electric potential ($\Delta V $) \citep{joun00,guichet03,guichet06,allegre10,allegrereply11}. The electrical potential $\zeta$  itself depends on fluid composition and $pH$, and the water conductivity \citep{davis78,ishido81,lorne99a,joun00,guichet06,jaafar09,vinogradov10,allegre10}.

\subsection{Packard's model}

\cite{packard53} proposed a model for the frequency-dependent streaming potential coefficient for capillary tubes,
assuming that the 
Debye length is negligible compared to the capillary radius,
 based on the Navier-Stokes equation:
\begin{equation}
\mathbf{C_{s0} (\omega)}= \frac{\Delta V(\omega)}{\Delta P(\omega)}= (\frac{\epsilon \zeta} {\eta \sigma_{f}}) (\frac{2}{a\sqrt{\frac{i\omega \rho_{f}}{\eta}}}  \frac{J_{1} (a\sqrt{\frac{i\omega \rho_{f}}{\eta}})} { J_{0} (a\sqrt{\frac{i\omega \rho_{f}}{\eta}})} e^{-i\omega t}) 
	\label{eqn:packard}
\end{equation}

where  $\omega$ is the angular frequency, $a$ is the capillary radius, $J_{1}$ and $J_{0}$ are the Bessel functions
 of the first order and the zeroth order, respectively,and $\rho_{f}$ is the fluid density.


The transition angular frequency for a capillary is:

\begin{equation}
\omega_{c}=\frac{\eta}{\rho_{f} a^{2}} 
\label{eqn:omegacpackard}
\end{equation}

More recently \cite{reppert01} used the low- and high-frequency approximations of the Bessel functions to propose the 
following formula, which corresponds to their eq.26 corrected with the right exponents $-2$ and $-1/2$:

\begin{equation}
\mathbf{C_{s0} (\omega )}= \left(\frac{\epsilon \zeta} {\eta \sigma_{f}}\right) \left[ 1+  \left(\frac{-2}{a} \sqrt{\frac{\eta}{\omega \rho_{f}}} \left(\frac{1} {\sqrt{2}}-\frac{1}{\sqrt{2}}i \right)\right)^{-2} \right]^{-\frac{1}{2}}
	\label{eqn:reppert}
\end{equation}
with the transition angular frequency
\begin{equation}
\omega_{c}=\frac{8\eta}{\rho_{f} a^{2}} 
\label{eqn:omegacreppert}
\end{equation}
and showed that this model was not very different from the model proposed by \cite{packard53}.


The complete development relating the Biot's theory and the Maxwell's equations has been published by Pride in 1994.

\subsection{Pride's model}

\cite{pride94} derived the equations governing the coupling between seismic and electromagnetic wave propagation in a fluid-saturated porous medium from first principles for porous media. The following transport equations express the coupling between the mechanical and electromagnetic wavefields [\citep{pride94} equations (174), (176), and (177)]:

\begin{equation}
	\mathbf{J}=\sigma (\omega) \mathbf{E} + L(\omega) \left( -\mathbf{\nabla} p + i \omega^{2} \rho_{f} \mathbf{u_{s}} \right)
	\label{eqn:DensiteCourant}
\end{equation}

\begin{equation}
	-i \omega \mathbf{w} = L(\omega) \mathbf{E} + \frac{k(\omega)}{\eta} \left( -\mathbf{\nabla} p+ i \omega^2 \rho_{f} \mathbf{u_{s}}\right)
	\label{eqn:MouvRelPoreFluide}
\end{equation}

In the first equation 
 the macroscopic electrical current density $\mathbf{J}$ is the sum of the average conduction and streaming current densities. The fluid flux $\mathbf{w}$ of the second equation 
 is separated into electrically and mechanically induced contributions. The electrical fields and mechanical forces that create the current density $\mathbf{J}$ and fluid flux $\mathbf{w}$ are, respectively, $\mathbf{E}$ and $(-\mathbf{\nabla} p +i \omega^{2} \rho_{f} \mathbf{u_s})$, where $p$ is the pore-fluid pressure, $\mathbf{u_s}$ is the solid displacement, and $\mathbf{E}$ is the electric field. 
The complex and frequency-dependent electrokinetic coupling $L(\omega)$, which describes the coupling between the seismic and electromagnetic fields \citep{pride94,reppert01} is the most important parameter in these equations. The other two coefficients, $\sigma (\omega)$ and $ k(\omega)$, are the electric conductivity and dynamic permeability of the porous material, respectively. 

The seismoelectric coupling that describes the coupling between the seismic and electromagnetic fields is complex and frequency-dependent \cite{pride94}:

\begin{equation}
	L(\omega) = L_{0} \left[1 - i \frac{\omega}{\omega_{c}} \frac{m}{4} \left(1 - 2\frac{d}{\Lambda} \right)^2  \left(1- i^{3/2} d \sqrt{\frac{\omega \rho_{f}}{\eta}} \right)^2 \right]^{-\frac{1}{2}}
	\label{eqn:CouplagePride}
\end{equation}
\noindent where $L_{0}$ is the low frequency electrokinetic coupling, $d$ is related to the Debye-length,
 $\Lambda$ is a porous-material geometry term \citep{johnson87}, and $m$ is a dimensionless number
 (detailed in \cite{pride94}).

The transition angular frequency $\omega_c$ separating low-frequency viscous flow and high-frequency 
inertial flow is defined as: 
\begin{equation}
\omega_{c}=\frac{\phi \eta}{\alpha_{\infty} k_{0} \rho_{f}} 
\label{eqn:omegacpride}
\end{equation}

\noindent where $\phi$ is the porosity, $k_0$ is the intrinsic permeability, $\alpha_{\infty}$ is the tortuosity.

\subsection{Further considerations}

The low-frequency electrokinetic coupling $L_{0}$ is related to the steady-state streaming potential coefficient 
$C_{s0}$ by:

\begin{equation}
 L_{0}= - C_{s0} \sigma_r
	\label{eqn:L0}
\end{equation}
where $\sigma_{r}$ is the rock conductivity.
The electrokinetic coupling $L(\omega)$ can be estimated by considering that steady-state models of $C_{s0}$ can be
 applied to the calculation of $L_{0}$. When writting $\sigma_{r}= \sigma_{f}/F$ with surface conductivity neglected,
 the steady-state electrokinetic coupling can be written as:
\begin{equation}
 L_{0}= - \frac{\epsilon \zeta}{\eta F}
	\label{eqn:L0F}
\end{equation}
 We can see that the steady-state electrokinetic coulping is inversely proportional to the formation factor.

The transition angular frequency separating viscous and inertial flows in porous medium can be rewritten by inserting
$ \alpha_{\infty}= \phi$ $ F $ with $F$ the formation factor that can be deduced from resistivity measurements
 using Archie's law, as:
 
\begin{equation}
\omega_{c} = \frac{1}{F} \frac{\eta}{ k_0 \rho_f}
\label{eqn:omegacprideKF}
\end{equation}

\noindent where $F$ is the formation factor that can be deduced from resistivity measurements using Archie's law.

Since the permeability and the formation factor are not independent, but can be related by 
$k_{0}= C R^{2} /F$ \citep{paterson83} with $C$ a geometrical constant usually in the range $0.3$-$0.5$ and $R$ the hydraulic radius, the transition angular frequency can be written as:
\begin{equation}
 \omega_{c} = \frac{\eta} {\rho_{f}C R^{2}} 
\label{eqn:omegacpaterson}
\end{equation}
The equation \ref{eqn:omegacpaterson} shows that the transition angular frequency in porous medium is inversely proportional to the square of the hydraulic radius.

Recently \cite{walker10} proposed a simplified equation of Pride's development assuming that the Debye length is negligible compared to the characteristic pore size, and assuming the parameter:

\begin{equation}
	m = 8 \left(\frac{\Lambda}{r_{eff}}\right)^{2}  
	\label{eqn:mWalker}
\end{equation}
leading to the equation:

\begin{equation}
	L(\omega) = L_{0} \left[1 - 2i \frac{\omega}{\omega_{c}} \left(\frac{\Lambda}{r_{eff}} \right) ^{2}  \right]^{-\frac{1}{2}}  
	\label{eqn:CouplageWalker}
\end{equation}

with $r_{eff}$  the effective pore radius, and a transition angular frequency 

\begin{equation}
\omega_{c}=\frac{8\eta}{\rho_{f} r_{eff}^{2}} 
\label{eqn:omegacwalker}
\end{equation}

\cite{garamb01} studied the low frequency assumption valid at seismic frequencies, meaning at frequencies 
lower than the Biot's frequency separating viscous and inertial flows and gave the coseismic transfer function 
for low frequency longitudinal plane waves. In this case, and assuming the Biot's moduli $C<<H$, they showed 
that the  seismoelectric field $\mathbf{E}$ is proportional to the grain acceleration:

 \begin{equation}
\mathbf{E}\simeq - \frac{L_{0}}{\sigma_{r}}\;\rho_{f} \mathbf{\ddot{u}} = \frac{\epsilon \zeta} {\eta \sigma_{f}}\;\rho_{f} \mathbf{\ddot{u}}
	\label{eqn:E}
\end{equation}

 Equations \ref{eqn:E}, \ref{eqn:L0} and \ref{eqn:Cs} show that transient seismo-electric magnitudes will be 
affected by the bulk density of the fluid, and the streaming potential coefficient which is inversely proportional to 
the water conductivity and proportional to the zeta potential (which depends on the water $pH$).

\subsection{The electrokinetic transition frequency compared to the hydraulic's one}

The theory of dynamic permeability in porous media has been studied by many authors \citep{auriault85,johnson87,sheng88,sheng88R,smeulders92}. 

The frequency behavior of the permeability is given by \cite{pride94} by:

\begin{equation}
	\frac{k(\omega)}{k_{0}} =  \left[ \left(1 - i \frac{\omega}{\omega_{c}} \frac{4}{m} \right)^{\frac{1}{2}} - i \frac{\omega}{\omega_{c}} \right]^{-1}
	\label{eqn:kdynamic}
\end{equation}

The transition angular frequency for a porous medium is the same as eq. \ref{eqn:omegacpride}. \cite{charlaix88} measured 
the behavior of permeability with frequency on capillary tube, glass beads and crushed glass. The dynamic permeability is
 constant up to the transition frequency above which it decreases, and the more permeable the sample is, the lower the 
transition frequency is. Other measurements have been performed on glass beads and sand grains \citep{smeulders92}. 
The transition frequency ($f_{c}=\omega_{c}/2 \pi$) varies from 4.8 Hz to 149 Hz for samples having permeability in 
the range $10^{-8}$ to $10^{-10}$ m$^{2}$ (see Table \ref{table:cutfrequency-glass}), which are extremely high permeabilities. 

The transition frequency indicates the beginning of the transition for both the permeability and the electrokinetic 
coupling. However the transition behavior and the cuttoff frequency are different between permeability and electrokinetic 
coupling (eq. \ref{eqn:CouplagePride} and 
eq.\ref{eqn:kdynamic}), both depending on the pore-space geometry term $m$ but in different manner.

\begin{figure}
  \centering
	\includegraphics[width=8cm]{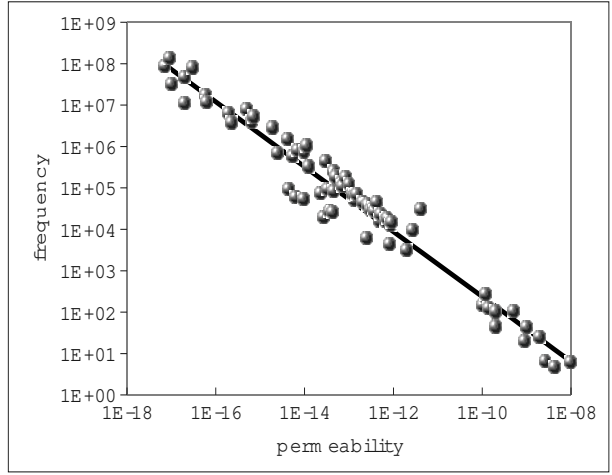}
	\caption{The transition frequency $f_{c}=\omega_{c}/2 \pi$ (in $Hz$) predicted in the present study with $\omega_{c}$ 
from eq. \ref{eqn:omegacprideKF} with $\eta = 10^{-3}$ Pa.s and $\rho_{f}= 10^{3}$ kg/m$^{3}$ as a function of 
the permeability (in $m^{2}$). The transition frequency varies as $log_{10} (f_{c}) = -0.78  log_{10}(k)-5.5$. The parameters of the samples, $F$ and $k_{0}$ are measured from different authors on various samples cited in Tables 1, 2 and 3}
	\label{fig:graphf-k}
\end{figure}

We calculated the predicted transition frequency $f_{c}=\omega_{c}/2 \pi$ with $\omega_{c}$ from eq. \ref{eqn:omegacprideKF} with $\eta = 10^{-3}$ Pa.s and $\rho_{f}= 10^{3}$ kg/m$^{3}$. The other parameters $F$ and $k_{0}$ are measured from different authors cited in \cite{bernabe91} (see Table \ref{table:cutfrequency}). We also calculated the parameters for four Fontainebleau sandstone samples. It has been shown for these samples that $F=\phi^{-2.01}$ (from \cite{ruffet91}) and that $k_{0}=a \phi^{n}$ with different values for $n$ according to the porosity. The following laws were chosen: $k_{0}=1.66$x$10^{-4}\phi^{8}$ for $\phi < 6\%$ and $k_{0}=2.5$x$10^{-10}\phi^{3}$ for $\phi$ ranging between $8$ and $25\%$ \citep{bourbie87}. We can see that the transition frequencies are of the order of kHz and MHz and no more from 0.2 to 150 Hz as measured or calculated on glass beads, sand grains, crushed glass or capillaries. We plotted the results of the transition frequency as a function of the permeability on these various samples in Fig. \ref{fig:graphf-k}. Although the formation factor is not constant with the permeability, it is clear that the transition frequency is inversely poportional to the permeability as:

\begin{equation}
 log_{10} (f_{c}) = -0.78  log_{10}(k)-5.5
\end{equation}

and varies from about $100$ MHz for $10^{-17}$ m$^{2}$ to about $10$ Hz for $10^{-8}$ m$^{2}$, so by seven orders of magnitude 
for nine orders of magnitude in permeability.


\section{Experimental apparatus and procedure}

Several experimental setups were proposed to provide the sinusoidal pressure variations.

\begin{figure}
  \centering
	\includegraphics[width=8cm]{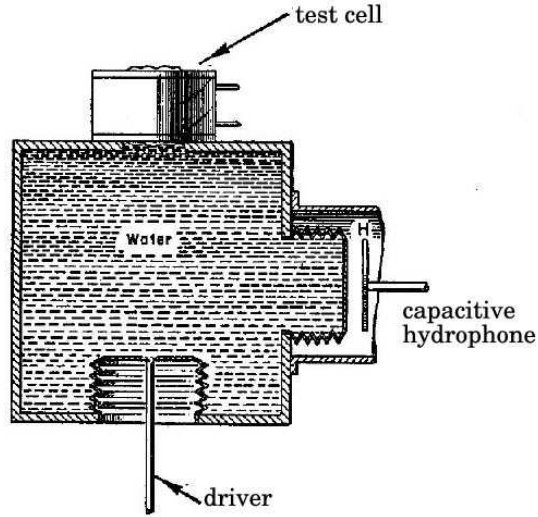}
	\caption{The sylphon bellows is driven by a geophone-type push-pull driver to apply a sinusoidal motion to the sample. (modified from \cite{packard53})}
	\label{fig:packard-exp}
\end{figure}

\begin{figure}
  \centering
	\includegraphics[width=8cm]{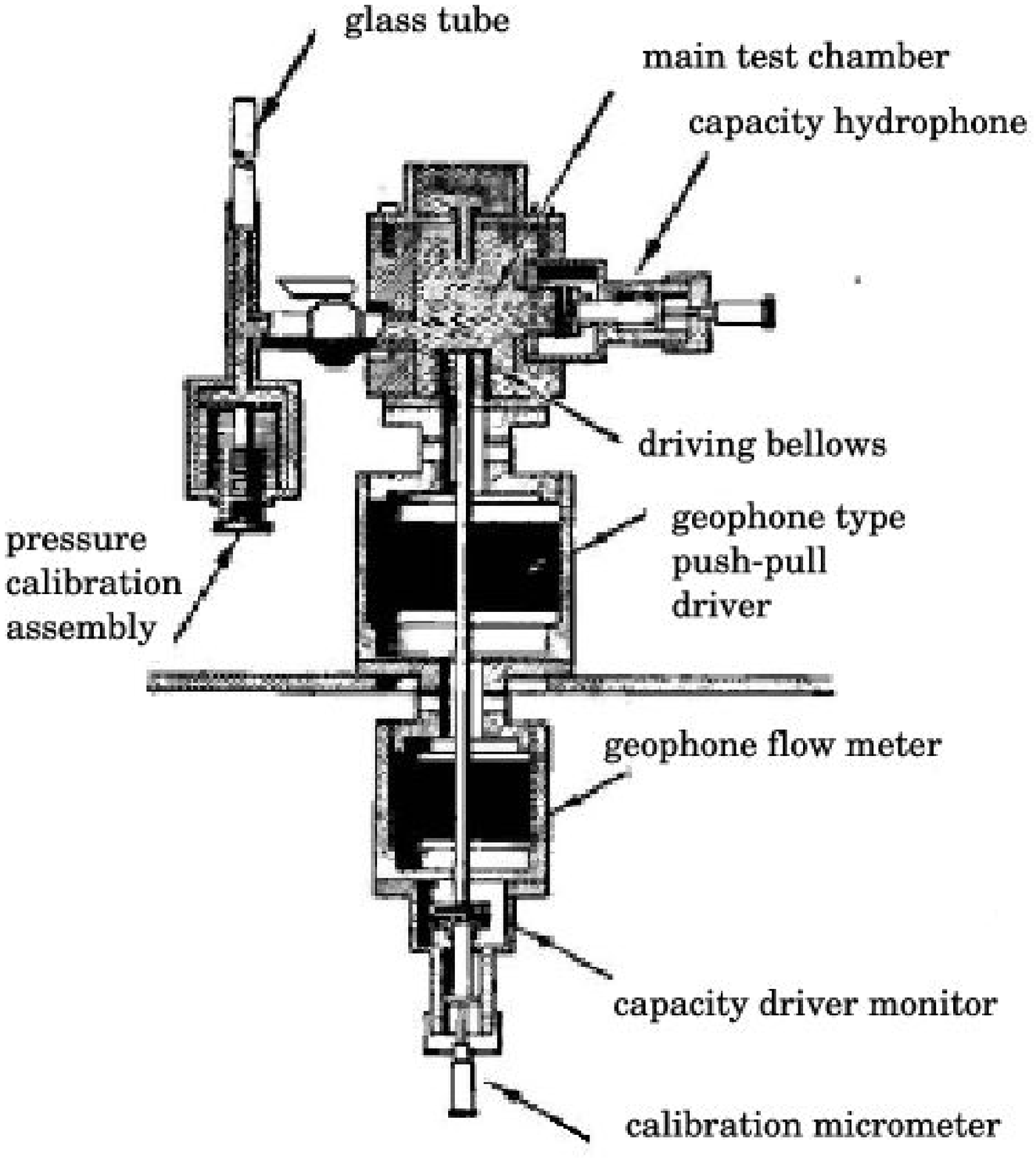}
	\caption{Experimental setup used by \cite{thurston52a} (modified from \cite{thurston52a}).}
	\label{fig:thurston-exp}
\end{figure}

\begin{figure}
  \centering
	\includegraphics[width=8cm]{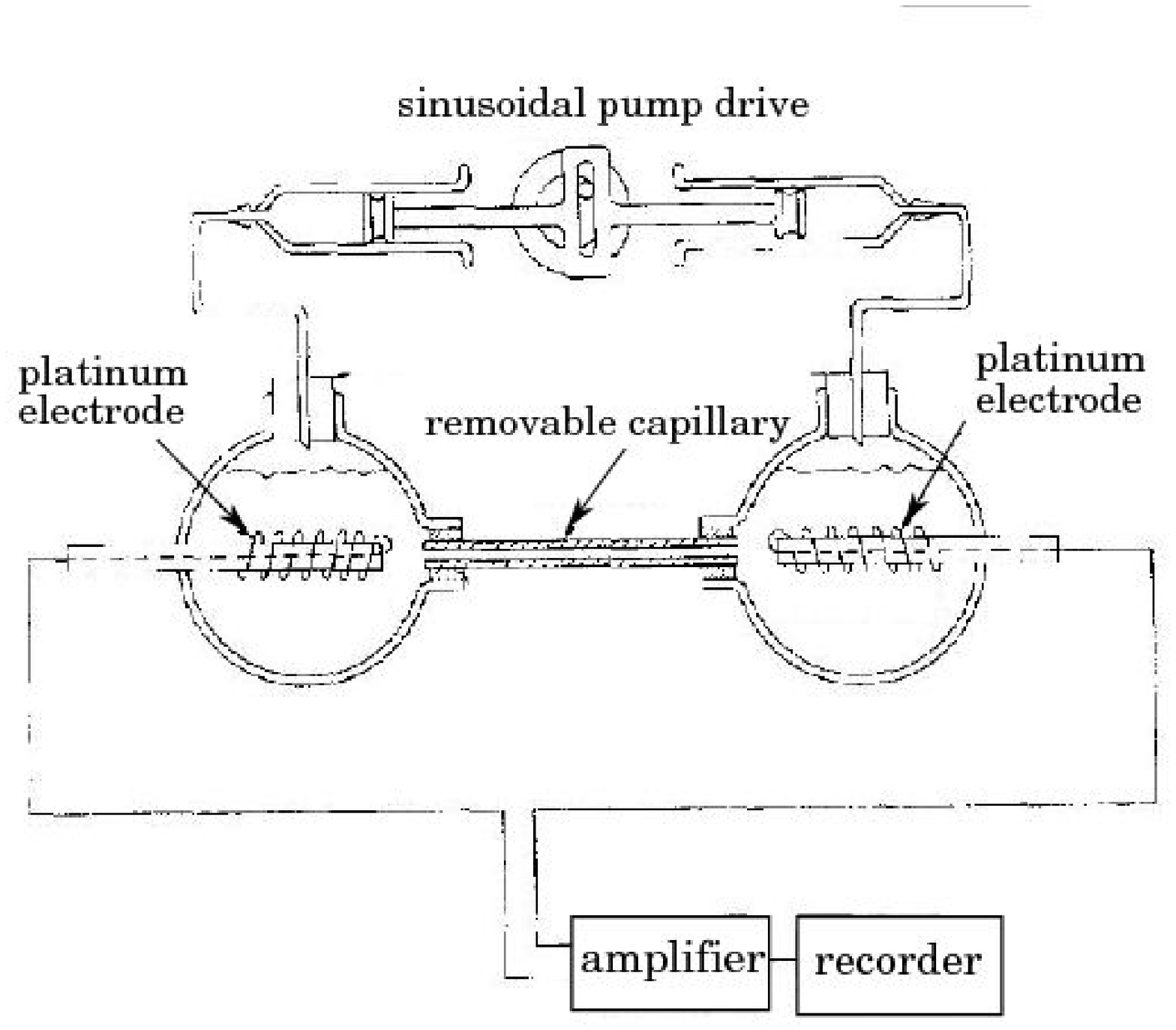}
	\caption{Experimental setup used by \cite{groves75} (modified from \cite{groves75}).}
	\label{fig:groves-exp}
\end{figure}

\begin{figure}
  \centering
	\includegraphics[width=8cm]{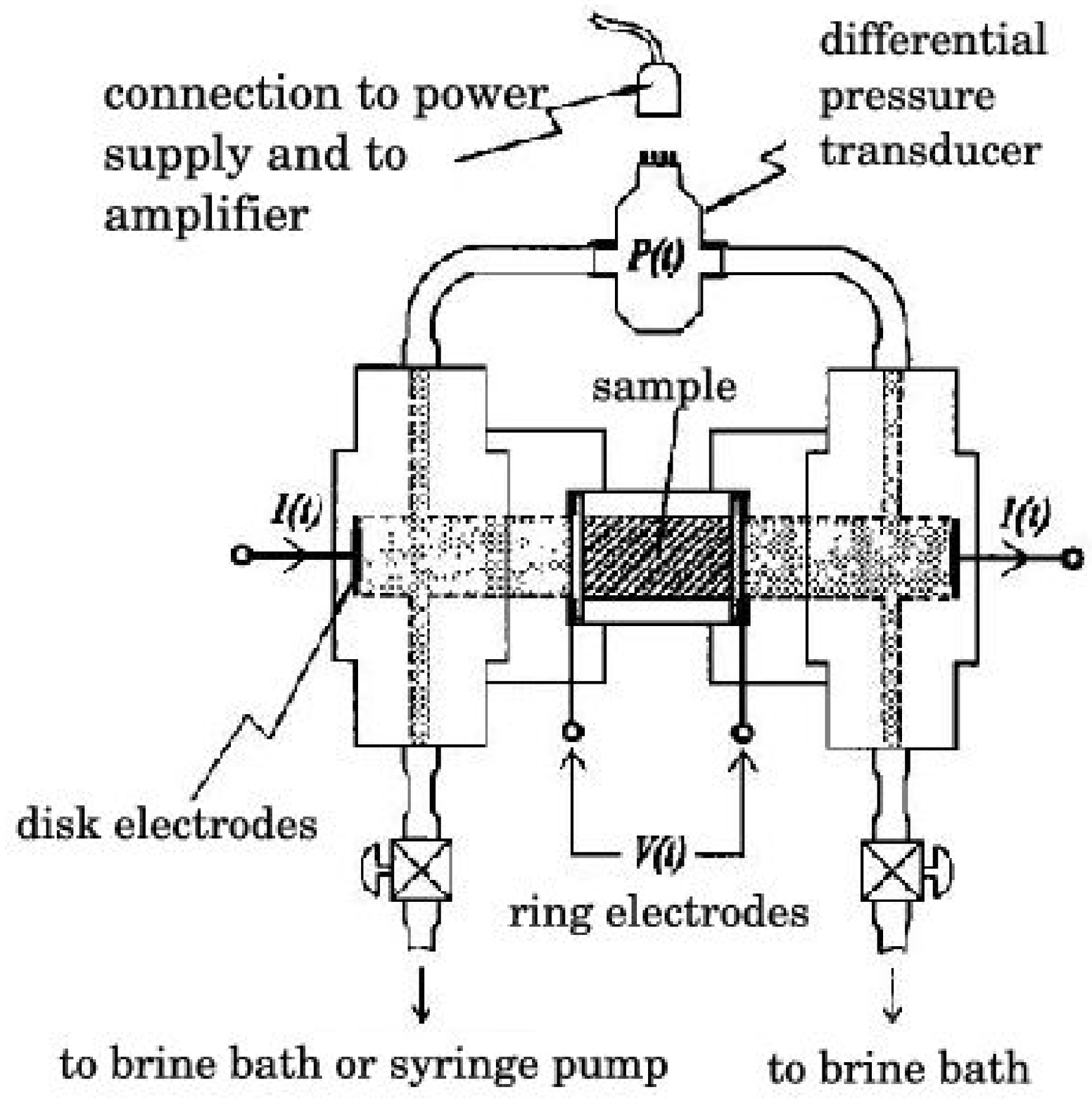}
	\caption{Experimental setup used by \cite{pengra99} for streaming potential and electro-osmosis measurements (modified from \cite{pengra99}).}
	\label{fig:pengra-exp}
\end{figure}

\begin{figure}
  \centering
	\includegraphics[width=8cm]{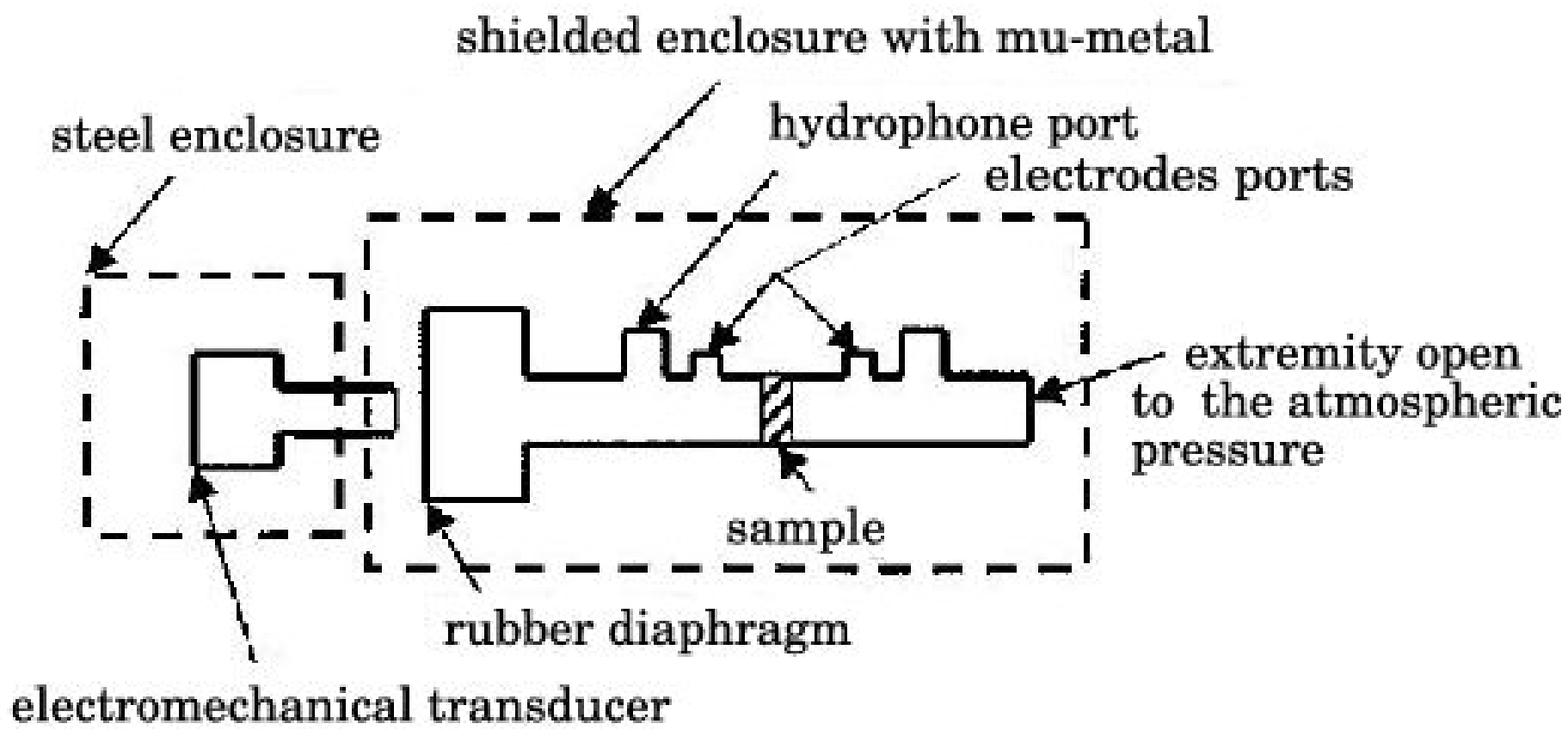}
	\caption{Experimental setup used by \cite{reppert01} (modified from \cite{reppert01}).}
	\label{fig:reppert-exp}
\end{figure}

The first experimental apparatus proposed a sinusoidal motion delivered by a sylphon bellows which was driven by a geophone-type push-pull driver (Fig. \ref{fig:packard-exp} from \cite{packard53}). The low frequency oscillator ($0.01$ Hz to $1$ kHz) was used for operation of the push-pull geophone driver. Similar setups were proposed by \cite{thurston52b} (Fig. \ref{fig:thurston-exp}) and \cite{cooke55}, so that frequency of this kind of source was $1$-$400$ Hz \citep{cooke55}, $20$-$200$ Hz \citep{packard53} and $10$-$700$Hz \citep{thurston52b}. The induced pressure was up to $2$ kPa. More recently \cite{schoemaker07} used a so-called Dynamic Darcy Cell (DCC) with a mechanical shaker connected to a rubber membrane leading to a frequency range for the oscillating pressure $5$ to $200$ Hz.
The sinusoidal fluid flow was also applied by a displacement piston pump directly connected to the electrodes chambers 
(fig. \ref{fig:groves-exp} from \cite{groves75,sears78}). The piston was mounted on a Scotch Yoke drive attached to a
 controllable speed AC motor \citep{cerda89}. The frequency range of this source was then $0.4$Hz to $21$ Hz and the pressure
 up to $15$ kPa. \cite{pengra99} used a piston rod attached to a loudspeaker driven by an audio power amplifier
 (Fig. \ref{fig:pengra-exp}). They performed measurements up to $100$Hz, with an applied pressure of $5$ kPa RMS. More recently
 it was proposed by \cite{reppert01} to use an electromechanical transducer (fig. \ref{fig:reppert-exp}), and these authors
 covered a frequency range $1$-$500$ Hz. The vibrating exciter proposed by \cite{schoemaker08} was used from $5$Hz to $200$Hz.
 Recently \cite{tardif11} used an electromagnetic shaker operating in the range $1$Hz to $1$kHz and provided measurements up
 to $200$Hz. Higher frequencies have been investigated \citep{zhu99,zhu00,chen05,block06,zhu08} for the detection of 
the interfacial conversions.

The electromagnetic noise radiating from such equipment must be suppressed by shielding the set-up and wires
 (shielded twisted cable pairs) \citep{tardif11,schoemaker08}. Moreover it is essential to have a rigid framework.
 A mechanical resonance can occur in the cell/transducer system (at $70$Hz in \cite{pengra99}), and the noise associated 
with mechanical vibration can be suppressed puting an additional mass to the frame \citep{tardif11}. 

Once the oscillatory pressure is applied, the pressure must be measured. Most of the setups include piezoelectric transducers to measure the pressure difference over the capillary or the porous sample. \cite{reppert01} proposed to use hydrophones that have a flat response from $1$ to $20$ kHz. \cite{tardif11} proposed to use dynamic transducers with a low-frequency limit $0.08$ Hz and a maximum frequency of $170$ kHz.

The electrodes are usually Ag/AgCl or platinium electrodes. The electrodes used by \cite{schoemaker08} were sintered plates of Monel (composed of nickel and copper).
The electrical signal must be measured using pre-amplifiers or a high-input impedance acquisition system. Since the impedance
 of the sample depends on the frequency, one must correct the measurements from this varying-impedance to be able to have a
 correct streaming potential coefficient \citep{reppert01}. Moreover the electrodes at top and bottom of the sample can behave
 as a capacitor, requiring a correction using impedance measurements too \citep{schoemaker08}.

The sample is usually saturated and it is emphasized that the sample should be left until equilibrium with water. This equilibrium can be obtained by leaving the sample in contact with water for some time, and by flowing the water within the sample several times by checking the $pH$ and the water conductivity until an equilibrium is reached \citep{guichet03}. The procedure including water flow is better because the properties of the water can be measured. When the properties of the water are measured only before saturating the sample, the resulting water once in contact with the sample is not known. Usually the water is more conductive when in contact with the sample, and the $pH$ can change. Recalling that the streaming potential is proportional to the zeta potential (which depends on $pH$) and inversely proportional to the water conductivity (eq.\ref{eqn:Cs}), it is essential to know properly the $pH$ and the water conductivity.


\section{Measurements and calculations of the dynamic electrokinetic coefficient}

\begin{figure}
  \centering
	\includegraphics[width=8cm]{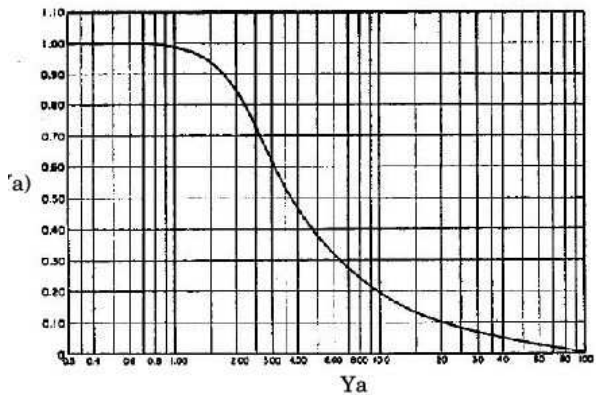}
	\caption{The absolute magnitude of the normalized streaming potential coefficient calculated by \cite{packard53} using eq. \ref{eqn:packardYa} where $Y_{a}= a \sqrt{\frac{\omega \rho_{f}}{\eta}}$, equivalent to eq. \ref{eqn:packard} (modified from \cite{packard53})}
	\label{fig:packard-modelYa}
\end{figure}

The absolute magnitude of the streaming potential coefficient normalized by the steady-state value was calculated
 by \cite{packard53} as:

\begin{equation}
f(Y_{a})= \left( \frac{-2}{Y_{a}} \frac{i \sqrt{i} J_{1} (\sqrt{i} Y_{a})} {J_{0} (\sqrt{i} Y_{a})} e^{-i\omega t}\right) 
	\label{eqn:packardYa}
\end{equation}

which is equal to eq. \ref{eqn:packard}, but expressed as a function of the parameter $Y_{a}= a \sqrt{\frac{\omega \rho_{f}}{\eta}}$, the transition frequency being obtained for $Y_{a}=1$ (Fig. \ref{fig:packard-modelYa}). The streaming potential coefficient is constant up to the transition angular frequency, and then decreases with increasing frequency.

\begin{figure}
  \centering
	\includegraphics[width=8cm]{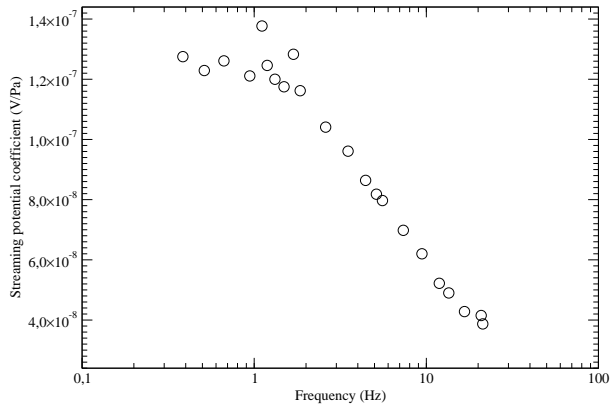}
	\caption{The streaming potential coefficient measured as a function of frequency by \cite{sears78} on a capillary coated with clay, incubated  with BSA in 0.02 M Tris-HCl.}
	\label{fig:sears78}
\end{figure}

\cite{sears78} measured the streaming potential coefficient on a capillary of radius $508$ $\micron$ which was coated with
 clay-Adams Siliclad and then incubated with 1\% bovine serum albumin, and filled with 0.02 $M$ Tris-HCl at $pH$ $7.32$. They 
reported the streaming potential and the pressure difference as a function of frequency in the range $0-20$ Hz. We calculated
 the resulting streaming potential coefficient (see Fig. \ref{fig:sears78}) which decreases from about $1.3$x $10^{-7}$ to 
$4$x $10^{-8}$ V/Pa. These authors computed the zeta potential and concluded that the zeta potential is independent of the 
frequency with an average value of $28.8$ mV. Moreover they concluded that the zeta potential is also independent of the 
capillary radius and capillary length.

The value of the streaming potential coefficient on Ottawa sand measured at 5 Hz by \cite{tardif11} was
 $-5.2$x $10^{-7}$ V/Pa using a 0.001 mol/L NaCl solution to saturate the sample. Values between $1$ and $2$x$10^{-8}$ V/Pa 
were measured on samples saturated by $0.1$ M/L NaCl brine \citep{pengra99}. A compilation of numerous streaming potential 
coefficients measured on sands and sandstones at various salinities in DC domain \citep{allegre10} showed that
 $C_{s0}=-1.2$ x $10^{-8} \sigma_f^{-1}$, where $C_{s0}$ is in V/Pa and $\sigma_{f}$ in S/m. A zeta potential of $-17$mV can
 be inferred from these collected data, assuming the other parameters (see eq. \ref{eqn:Cs}) independent of water conductivity. 
These assumptions are not exact, but the value of zeta is needed for numerous modellings which usually assume the other
 parameters independent of the fluid conductivity. Therefore an average value of $-17$ mV for such modellings can be rather exact, at least for medium with no clay nor calcite.

\begin{figure}
  \centering
	\includegraphics[width=8cm]{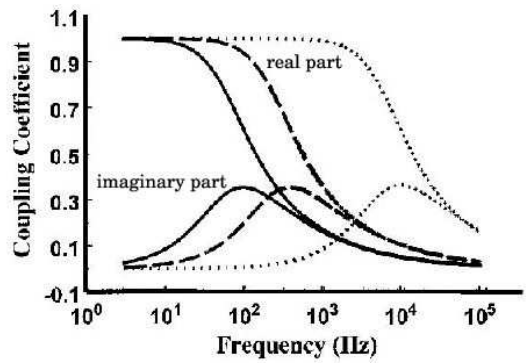}
	\caption{The real and imaginary part of the Packard's model (eq.\ref{eqn:packard}) calculated by \cite{reppert01} for three capillary radii: 100\micron (continuous line), 50\micron (dashed line), 10\micron (point line) (modified from \cite{reppert01}).}
	\label{fig:reppert-modelP}
\end{figure}

 \cite{reppert01} calculated the real part and the imaginary part of the theoretical Packard's streaming potential coefficient (eq. \ref{eqn:packard}) for different capillary radii. (see Fig. \ref{fig:reppert-modelP}). It can be seen that the larger the radius is, the lower the transition frequency is, as shown above by the different theories. Recent developments by the group of Glover have been performed to build a new setup and to make further measurements on porous samples: two papers detail these studies in this special issue on Electrokinetics in Earth Sciences.

\section{Conclusion}
Since the theory of Pride in 1994, the dynamic behavior of the streaming potential is known for porous media. However few 
experimental results are avalaible, because of the difficulty to perform correct measurements at high frequency. Up to now,
 measurements of the frequency-dependence of the streaming potential have been performed up to $200$ Hz on high-permeable samples.
 The main difficulty arises from electrical noise induced by mechanical vibration. Moreover it has been emphasized that the measurements must be corrected by impedance measurements as a function of frequency too because the impedance of the sample depends on frequency. Further theoretical developments performed by \cite{garamb01} studied the low frequency assumption valid at frequencies lower than the transition frequency. We show that this transition frequency, on a various collection of samples for which both formation factor and permeability are measured, is predicted to depend on the permeability as inversely proportional to the permeability. 

\section{Acknowledgements}
This work was supported by the French National Scientific Center (CNRS), by the National Agency for Research (ANR) 
through TRANSEK, and by REALISE the ``Alsace Region Research Network in Environmental Sciences in Engineering'' and the 
Alsace Region. We thank two anonymous reviewers and the associate editor T. Ishido for very constructive remarks that 
improved this paper.

\section{References}

\bibliographystyle{gji}
\bibliography{biblio}


\newpage

\begin{table*}
\caption{Measured or predicted transition frequency for dynamic streaming potential and permeability, for samples of porosity $\phi$, formation factor $F$, permeability $k_{0}$, and half of the mean particle size r, from (SED) \cite{smeulders92}, (CKS) \cite{charlaix88}, (SG) \cite{sears78}, (P)\cite{packard53}, (TGR) \cite{tardif11}, (RMLJ) \cite{reppert01}. $^{*}$ indicates predicted transition frequency from eq. \ref{eqn:omegacpackard} and $^{**}$ indicates the transition frequency computed by the authors.}
\label{table:cutfrequency-glass}
\begin{tabular}{@{}|c|c|c|c|c|c|c|c|c|c|}
Sample & particle size $\mu m$ & $\phi $ [\%] &$F$ & $k_{0}$ [m$^{2}$] & $f_{c}$ [Hz] & source \\
\hline
capillary & 254(radius) &  &  & 10$^{-8}$ & 10-2.5$^{*}$ Hz & CKS\\
capillary & 508(radius) &  &  &  & 1.3-0.62$^{*}$ Hz & SG\\
capillary G4 & 720(radius) &  &  & & 0.31$^{*}$-0.28 $^{**}$ Hz & P\\
capillary G2 & 826(radius) &  &  & & 0.23$^{*}$-0.21 $^{**}$ Hz & P\\
capillary 1 & 800-1100(radius) &  &  & & 7.1 Hz & RMLJ\\
glass beads & 1.25-1.75 & 32 & 7.8 & 4.2x10$^{-9}$ & 4.8 Hz & SED\\
glass beads & 850 (r) & 50 & 2.8 & 10$^{-8}$ & 6.2 Hz & CKS\\
glass beads & 580-700 & 31 & 8.7 & 9x10$^{-10}$ & 20 Hz & SED\\
glass beads & 450 (r) & 50 & 3.2 & 2x10$^{-9}$ & 25 Hz & CKS\\
glass beads & 250 (r) & 50 & 3 & 5x10$^{-10}$ & 108 Hz & CKS\\
glass beads & 200-270 & 31 & 9 & 1.4x10$^{-10}$ & 126 Hz & SED\\
crushed glass & 440 (r) & 50 & 3 & 10$^{-9}$ & 44 Hz & CKS\\
crushed glass & 265 (r) & 50 & 3.2 & 2x10$^{-10}$ & 45-103 Hz & CKS\\
porous fliter A & 72.5-87 &  &  &  & 269 Hz & RMLJ\\
porous fliter B & 35-50 &  &  &  & 710 Hz & RMLJ\\
sand grains & 1000-2000 & 31 & 9 & 26x10$^{-10}$ & 6.7 Hz & SED\\
sand grains & 150-300 & 29 & 10.7 & 10$^{-10}$ & 149 Hz & SED\\
Ottawa sand  & 200-250 (r) & 31 & 4.7 & 1.2x10$^{-10}$ & 230-273 Hz & TGR\\
\end{tabular}
\end{table*}

\newpage

\begin{table*}
\caption{Predicted transition frequency (from eq. \ref{eqn:omegacprideKF}) for dynamic streaming potential, for samples of porosity $\phi$, formation factor $F$ and permeability $k_{0}$, from (1) calculated in the present study, and measured by (2) \cite{taherian90}, (3) \cite{morgan90}, (4)\cite{fatt57}, (5)\cite{wyble58}, (6)\cite{dobrynin62}, (7)\cite{chierici67}, (8)\cite{yale84}.}
\label{table:cutfrequency}
\begin{tabular}{@{}|c|c|c|c|c|c|c|c|c|c|}
Sample & $\phi $ [\%] &$F$ & $k_{0}$ [m$^{2}$] & $f_{c}$ [Hz] \\
\hline
Fontainebleau sandstone$^{1}$ & 20 & 25 & 2x10$^{-12}$ & 3.2 kHz \\
Fontainebleau sandstone$^{1}$ & 15 & 45 & 8x10$^{-13}$ & 4.4 kHz \\ 
Fontainebleau sandstone$^{1}$ & 10 & 102 & 2.5x10$^{-13}$ & 6.2 kHz \\ 
sandstone-S22$^{2}$ & 31.2 & 6 & 2.7x10$^{-12}$ & 9.7 kHz \\ 
sandstone-S47$^{2}$ & 20 & 14.4 & 8.5x10$^{-13}$ & 13 kHz \\ 
Boise$^{8}$ & 26 & 12 & 9x10$^{-13}$ & 14.7 kHz \\
Berea sandstone500$^{8}$ & 20 & 20 & 4.9x10$^{-13}$ & 16.2 kHz \\ 
sandstone-S42$^{2}$ & 19.7 & 14.7 & 6.7x10$^{-13}$ & 16.2 kHz \\ 
sandstone-S45$^{2}$ & 21 & 11.7 & 7.2x10$^{-13}$ & 18.8 kHz \\ 
Fahler 162$^{8}$ & 3 & 294 & 2.7x10$^{-14}$ & 20 kHz \\
sandstone-S43$^{2}$ & 21.2 & 13 & 5.1x10$^{-13}$ & 23.5 kHz \\ 
Pliocene 41$^{7}$ & 21 & 144.9 & 4.2x10$^{-14}$ & 26.1 kHz \\ 
Pliocene 35$^{7}$ & 20 & 156.2 & 3.7x10$^{-14}$ & 27.5 kHz \\ 
Berea sandstoneC2H$^{3}$ & 19.8 & 15.1 & 3.8x10$^{-13}$ & 27.7 kHz \\ 
sandstone-S50$^{2}$ & 18.3 & 17.2 & 3.1x10$^{-13}$ & 30 kHz \\ 
Triassic38$^{7}$ & 21 & 12.6 & 4x10$^{-12}$ & 31.4 kHz \\ 
Triassic34$^{7}$ & 20 & 13.9 & 3.5x10$^{-13}$ & 32.7 kHz \\ 
Berea sandstoneB2$^{3}$ & 20.3 & 15.2 & 2.64x10$^{-13}$ & 39.7 kHz \\ 
sandstone-S5$^{2}$ & 26.4 & 8.7 & 4.1x10$^{-13}$ & 45 kHz \\ 
sandstone-S35$^{2}$ & 18.75 & 17.4 & 2x10$^{-13}$ & 46.5 kHz \\ 
Massillon DH$^{8}$ & 16 & 23.8 & 1.3x10$^{-13}$ & 51.4 kHz \\
Cambrian 16$^{7}$ & 14 & 312.5 & 9.5x10$^{-15}$ & 53.6 kHz \\ 
Fontainebleau sandstone$^{1}$ & 5 & 412 & 6.5x10$^{-15}$ & 59.4 kHz \\ 
Berea sandstoneD1$^{3}$ & 18.5 & 18.4 & 1.3x10$^{-13}$ & 66.5 kHz \\ 
Tensleep1$^{4}$ & 15 & 18.9 & 1.2x10$^{-13}$ & 70.3 kHz \\ 
Tertiary 807$^{8}$ & 22 & 14.9 & 1.5x10$^{-13}$ & 71.1 kHz \\
Cambrian 6$^{7}$ & 8.1 & 90.9 & 2.3x10$^{-14}$ & 76.1 kHz \\ 
\end{tabular}
\end{table*}

\begin{table*}
\caption{continued}
\label{table:cutfrequency2}
\begin{tabular}{@{}|c|c|c|c|c|c|c|c|c|c|}
Sample & $\phi $ [\%] &$F$ & $k_{0}$ [m$^{2}$] & $f_{c}$ [Hz] \\
\hline
Torpedo$^{6}$ & 20 & 41.7 & 4.5x10$^{-14}$ & 84.9 kHz \\ 
Miocene 7$^{7}$ & 8.3 & 384.6 & 4.4x10$^{-15}$ & 94 kHz \\ 
Cambrian 14$^{7}$ & 11 & 52.6 & 3.2x10$^{-14}$ & 94.5 kHz \\ 
sandstone Triassic27$^{7}$ & 18 & 20 & 7.2x10$^{-14}$ & 110.5 kHz \\ 
sandstone-S9$^{2}$ & 20.9 & 12 & 1x10$^{-13}$ & 126.2 kHz \\ 
Triassic26$^{7}$ & 18 & 17.2 & 6.8x10$^{-14}$ & 135.7 kHz \\ 
sandstone-S6$^{2}$ & 22.8 & 10.6 & 8.3x10$^{-14}$ & 180.7 kHz \\ 
Berea 100H$^{8}$ & 17 & 17.2 & 4.9x10$^{-14}$ & 188.4 kHz \\
sandstone S15$^{2}$ & 21.8 & 13.9 & 4.5x10$^{-14}$ & 256.7 kHz \\ 
Kirkwood$^{5}$ & 15 & 40 & 1.2x10$^{-14}$ & 331.6 kHz \\ 
Indiana DV$^{8}$ & 27 & 12 & 3x10$^{-14}$ & 440.3 kHz \\ 
Island Rust A1$^{3}$ & 14.6 & 52.5 & 5.2x10$^{-15}$ & 579 kHz \\ 
Bradford$^{5}$ & 11 & 90 & 2.5x10$^{-15}$ & 700.3 kHz \\ 
Austin chalk$^{3}$ & 23.6 & 22.7 & 9.7x10$^{-15}$ & 763 kHz \\ 
Massillon DV$^{8}$ & 19 & 27.8 & 6.9x10$^{-15}$ & 830.4 kHz \\
sandstone-S34$^{2}$ & 21.35 & 13.7 & 1.1x10$^{-14}$ & 1.06 MHz \\ 
sandstone S44$^{2}$ & 15.7 & 24.5 & 4.2x10$^{-15}$ & 1.5 MHz \\ 
Indiana L. SA1$^{3}$ & 18 & 29.2 & 1.9x10$^{-15}$ & 2.9 MHz \\ 
Tennessee A1$^{3}$ & 5.5 & 180.3 & 2.3x10$^{-16}$ & 3.8 MHz \\ 
AZPink (Coconino)$^{3}$ & 10.3 & 62.4 & 6.3x10$^{-16}$ & 4.04 MHz \\ 
Leuders L.SA1$^{3}$ & 15.2 & 41.5 & 7.1x10$^{-16}$ & 5.3 MHz \\ 
sandstone-S40$^{2}$ & 10.9 & 130 & 1.9x10$^{-16}$ & 6.4 MHz \\ 
sandstone-S23$^{2}$ & 18.8 & 40.7 & 4.8x10$^{-16}$ & 8.1 MHz \\ 
Fahler 189$^{8}$ & 1.9 & 714.3 & 2x10$^{-17}$ & 11.1 MHz \\
Penn blue A1$^{3}$ & 3.9 & 219 & 6.2x10$^{-17}$ & 11.7 MHz \\
AZChoclate2$^{3}$ & 9.5 & 159.3 & 5.8x10$^{-17}$ & 17.2 MHz \\
Fahler 161$^{8}$ & 2.3 & 416.7 & 1x10$^{-17}$ & 38.2 MHz \\
Fahler 142$^{8}$ & 7.6 & 164 & 2x10$^{-17}$ & 48.5 MHz \\
sandstone S21$^{2}$ & 12.1 & 65 & 3x10$^{-17}$ & 81.7 MHz \\
Fahler 154$^{8}$ & 4.6 & 263.1 & 7x10$^{-18}$ & 86.4 MHz \\
Fahler 192$^{8}$ & 4.4 & 128.2 & 9x10$^{-18}$ & 137.9 MHz \\
\end{tabular}
\end{table*}

\newpage

\end{document}